\documentclass[11pt]{article}

\begin{document}
\pagestyle{plain}
\begin{center}
\vskip10mm
   {\Large {\bf Structure behind Mechanics: Overview}}
\vskip10mm
\renewcommand{\thefootnote}{\dag}
{\large Toshihiko Ono}\footnote{ 
e-mail: BYQ02423@nifty.ne.jp \ \ or \ \ 
tono@swift.phys.s.u-tokyo.ac.jp}
\vskip5mm
\par\noindent
{\it 703 Shuwa Daiich Hachioji Residence,\\
4-2-7 Myojin-cho, Hachioji-shi, Tokyo 192-0046, Japan}
\vskip5mm
\end{center}
\vskip10mm
\setcounter{footnote}{0}
\renewcommand{\thefootnote}{\arabic{footnote}}

\begin{abstract}
This letter proposes
a new scenario to solve the structural or conceptual
problems
remained in quantum mechanics, and 
gives an overview of the 
theory proposed in
quant-ph/9906130 (including quant-ph/9909025 and 
quant-ph/0001015).


\begin{center}
{\it Submitted to Physics Letters A.}
\end{center}
\end{abstract}
\vskip10mm

Although quantum mechanics in twentieth century
has succeeded to predict 
the correct results of large numbers of experiments
in the process to find new fundamental particles
in the nature,
it
seems to have left some fundamental open problems,
structural or conceptual ones:
\begin{enumerate}
\item the structural problems of
\begin{itemize}
\item the operator ordering \cite{Groenwald, van Hove},
\item the analyticity at the exact classical-limit of $\hbar =0 $
\cite{Maslov,Truman} and
\item the semantics of the reguralization in quantum field 
theories
\cite{Pauli&Villars,tHooft&Vertman};
\end{itemize}
\item the conceptual problems of
\begin{itemize}
\item the universal validity \cite{Mittelstaedt},
\item the wave-reduction in measurement processes 
\cite{BLM} and
\item the compatibility with causality \cite{EPR, Bell, 
Aspect}.
\end{itemize}
\end{enumerate}
These difficulties come from the problem
how and why quantum mechanics relates itself with 
classical mechanics:
the relationship between the quantization that
constructs quantum mechanics based on
classical mechanics
and the classical-limit
that induces classical mechanics
from quantum mechanics as an approximation
with Planck's constant $h$ taken to be zero;
the incompatibility between
the ontological feature
of classical mechanics and 
the epistemological feature
of quantum mechanics
in commonly accepted interpretations.
The present theory,\footnote{
Almost all the information
of the introduced theory 
has appeared in the newest version of quant-ph/9906130 
on the LANL preprint server.}
originated by the previous letter \cite{Ono},\footnote{ 
The author 
of letter \cite{Ono}, "Tosch Ono," 
is the same person as that of the present paper,
"Toshihiko Ono."} aims
to solve {\it all} of the above listed problems in quantum 
mechanics.
Let me here present just an overview of the theory to 
inform 
the vital conclusions obtained from the theory {\it 
without} the
mathematical and/or semantic complexities,
while detail descriptions will be made in a series of full 
papers \cite{I}.

The
proposed theory on 
physical reality, 
named as {\it Structure behind Mechanics} (SbM),
 supposes that a field or a particle $X$
on the four-dimensional spacetime
has its internal-time $\tilde o_{{\cal P}(t)}(X)\in S^1 $ 
relative
to a domain ${\cal P}(t)$ 
of the four-dimensional spacetime,
whose boundary and interior
represent the present and the past at ordinary time $t$, 
respectively.
For $\bar h = \hbar /2$,
the classical action $\bar h S_{{\cal P}(t)}(X)\in {\bf R}$ 
realizes
internal-time $\tilde o_{{\cal P}(t)}(X) $ in the following 
relation:
\begin{equation}
\tilde o_{{\cal P}(t)}(X) = e^{iS_{{\cal P}(t)}(X)}.
\end{equation}
Object $X$ also has the 
external-time $\tilde o_{{\cal P}(t)}^*(X)\in S^1 $ 
relative
to  ${{\cal P}(t)}$  which is the internal-time of all the 
rest but
$X$ in the universe.
It gains the actual existence on  ${{\cal P}(t)}$
if 
and only if the internal-time
coincides with the external-time in a quasi-periodic way:
\begin{equation}\label{AX=XA}
\tilde o _{{\cal P}(t)}(X)= \tilde o_{{\cal P}(t)}^*(X) .
\end{equation}
This condition discretizes or quantizes the ordinary time 
passing from 
the past to the future,
and mathematically supports
Whitehead's epochal theory of time \cite{Whitehead}.
It also shows that 
object $X$ has its actual existence
only when it is 
exposed to 
or has the possibility
to interact with the rest of the world,
and illustrates 
that
such existence
can become the empirical one
through the actual interaction with the open system
as in a measurement process.
The indeterministic 
influence  of the outside  on object $X$ 
would cause the irreversibility of time
on the fundamental level.
In this way,
the present theory can
presuppose that there exist three
levels of existence:
\begin{enumerate}
\item the ideal existence: the immutable being or 
potentiality,
\item the actual existence: the becoming or emergence,
\item the empirical existence: the appearing or detected.
\end{enumerate}
The ordinary dogma of
quantum mechanics has admitted only
the third level of the indeterministic existence,
while the deterministic realism in classical mechanics
has accepted the first category.
The present theory
considers that both mechanics'
really refer 
the second kind of
existence, {\it emergence},
which can be defined here as deterministic process from 
the inside in terms of the ideal existence
and that must be described later as indeterministic 
process from the outside including an observer 
in terms of the empirical existence. 
As such, it reconciles two different ideas of reality
in
general relativity and quantum theory;
it also provides
the regularization method in quantum field theories
with the semantics
that a regularization parameter
is corresponding to the time-interval
of the emergences of a particle.
The both sides of relation (\ref{AX=XA}) further
obey the variational principle as
\begin{equation}
 \label{dAX=0/dXA=0}
\delta \tilde o_{{\cal P}(t)} (X)   = 0
\ \ \ , \ \ \ \
\delta \tilde o^*_{{\cal P}(t)} (X) = 0 .
\end{equation}  
These equations produce the equations of motion
in the deduced mechanics.

Theory of SbM, the description of a system 
from inside,
provides a foundation
for quantum mechanics and classical mechanics
as the description of a system from outside,
named as 
{\it protomechanics}.\footnote{Please 
allow me
to name the proposed theory
in this way
for convenience 
tentatively. It can 
not be regarded, of course, as a new 
mechanics in a 
physical sense until it suffers the 
several experimental 
and theoretical tests.}
The space $M$ of all the objects over present 
hypersurface 
$\partial {\cal P}(t)$
have an mapping $o_t : TM \to S^1$ for
the position $(x_t, \dot x_t )$ in the cotangent space
$TM$ 
corresponding to an object $X$:
\begin{equation}\label{def eta0}
o_t \left( x_t,  \dot x_t \right) = 
\tilde o_{{\cal P}(t)} \left( X \right)  .
\end{equation}
For the velocity field $v_t $ over $M$ 
such that $v_t \left( x_t\right) = 
 {{dx_t}\over {dt}} $, we will introduce a
section $\eta_t $
and call it {\it synchronicity} over $M$:
\begin{equation}\label{def eta}
\eta_t (x)=o_t \left( x, v_t (x) \right)   ;
\end{equation}
thereby, synchronicity $\eta_t$
has an 
information-theoretical sense,
as
defined for
a collective set of objects
that have different initial conditions
from one another.
On the other hand,
the emergence-frequency
$f_t\left( \eta_t \right) $  represents
the frequency that
object $ X$ satisfies condition (\ref{AX=XA}) on $ M$;
and
the true probability measure $\nu_t $ on $TM$,
 representing the ignorance of the initial
position,
 defines the {\it emergence-measure}
 $\mu_t \left( \eta_t \right) $ as follows:
\begin{equation}
d\mu_t \left( \eta_t \right) (x)= d\nu_t \left( x, v_t(x) 
\right)  \cdot 
f_t
\left( \eta_t \right)  (x)  .
\end{equation}
Through a measurement
process,
the above defined emergence-measure 
becomes the probability measure for the detection of a
particle if positive everywhere.
The emergence-measure for the observables measured 
in indirect ways can partially have negative values 
since there are two cases that internal-time
 $\tilde o_t(X)$ exceed  $\tilde o^*_t(X)$ for condition 
(\ref{AX=XA})
 and viceversa. Such negativity partially causes the 
strange
 behaviors of non-commuting observables \cite{Moyal} 
and
 the breaking of Bell's inequality \cite{Bell}. 
The induced Hamiltonian $H^{T^*M}_t$ on $T^*M$,
further,
redefines the velocity field $v_t$
and the Lagrangian $L^{TM}_t$ as follows:
\begin{eqnarray}
v_t (x)
&=& {{\partial H^{T^*M}_t}\over {\partial p}}\left( x,
p\left( \eta_t \right) (x)
\right) \\
L^{TM}_t \left( x,
v(x)  \right)
&=&  v  (x) \cdot p \left(\eta _t \right) (x)   -  
H^{T^*M}_t\left( x, p\left(\eta _t \right) (x)  \right) ,
\end{eqnarray}
where mapping
$p$ satisfies
the modified Einstein-de Broglie relation:
\begin{equation}
p \left(\eta _t \right) =
- i \bar h   \eta_t  ^{ -1}
d \eta_t .
\end{equation}
The equation of motion
is the set of the following equations:
\begin{eqnarray}\label{EM1}
\left( {{\partial }\over {\partial t}}
+{\cal L}_{v_t} \right) \eta_t (x)&=&  -i {\bar h}^{-1}
L_t^{TM}\left( x, v_t (x)\right)   \eta_t (x) ,\\
\label{EM2}
\left( {{\partial }\over {\partial t}}
+{\cal L}_{v_t} \right) d\mu_t \left( \eta_t \right) &=& 0.
\end{eqnarray}

Protomechanics has the statistical description of
the set
$\Gamma $ of all the
synchronicities on space $M$.
To investigate such a description,
we will introduce the related group.
The group ${\cal D}(M)$ of
all the
$C^{\infty }$-diffeomorphisms of $M$
and the abelian group
$ C^{\infty }\left( 
M \right) $
 of all the $C^{\infty }$-functions 
on $M$
construct the semidirect product
$ S  ( M ) =
{\cal D} (M) \times_{semi. } C^{\infty }(M) $,
and define
the multiplication $\cdot $
between $ \Phi_1=(\varphi_1,s_1)$ and $\Phi_2=(\varphi_2,s_2)\in   S  ( M  )
$ as  
\begin{equation}\label{(2.1.1)}
\Phi_1\cdot \Phi_2 =(\varphi _1 \circ \varphi 
_2,(\varphi_2^*
s_1 )\cdot s_2) ,  
\end{equation}
for the pullback $\varphi ^*$ by $\varphi \in
{\cal D} (M)$ (consult \cite{M&R&W}).
We shall further introduce the group $Q(M) 
= Map\left( \Gamma , S  (M) \right) $ 
of all the mapping from $ \Gamma  $ into $S(M) $,
that has
the Lie algebra $q(M) 
= Map\left(  \Gamma  , s  (M)\right) $ and
its dual space $ 
q(M)^* = Map \left(   \Gamma  , s  (M)^* \right)  $.
Synchronicity $\eta ^{\tau }_{t }(\eta )$,
such that
 the {\it labeling time} $\tau $ satisfies
$\eta ^{\tau }_{\tau }(\eta ) =\eta  $,
has the momentum
 $p_t^{\tau }(\eta )
=-i \bar h\eta ^{\tau }_t(\eta )^{-1}
d \eta ^{\tau }_t(\eta ) $ and
the emergence-measure $\mu^{\tau }_t(\eta ) $
such that
\begin{eqnarray}\label{mu(F)}
\bar  \mu_t \left( p^*F_t\right) 
&=&\int_{\Gamma }d {\cal M} (\eta ) \
 \tilde \mu_t (\eta ) \ \left( p^*F_t  (\eta )
  \right)  \\
&=& \int_{\Gamma }d {\cal M} \left( \eta \right) \
\tilde \mu^{\tau }_t \left( \eta  \right) \ \left( p^*F  \left( 
\eta ^{\tau }_t(\eta )\right) 
  \right)  ,
\end{eqnarray}
where ${\cal M}$ is a probability measure on 
$\Gamma $.
The introduced labeling time $\tau $
can always be chosen such that $ \eta ^{\tau }_t(\eta ) $
does not have any singularity
within a short time for every $\eta $.
The emergence-momentum
$ {\cal J}_t^{\tau }  
\in   q\left(M\right) ^*$ 
such that
\begin{eqnarray}
{\cal J}_t^{\tau }  (\eta )=  d {\cal M} (\eta ) \
 \left(  \tilde \mu_t ^{\tau }
\left( \eta  \right) \otimes 
p_t^{\tau }(\eta )  ,
\tilde \mu_t ^{\tau }
\left( \eta  \right) \right) \ 
\end{eqnarray}
satisfies the following relation
for the functional
${\cal F}_t   : q\left(M\right) ^* \to {\bf R} $:
\begin{equation}
{\cal F}_t  \left(  {\cal J}^{\tau }_t\right)
=\bar \mu_t \left( p^* F_t\right) ,
\end{equation}
whose value is independent of
labeling time $\tau $.
For Hamiltonian operator $\hat H_t^{\tau }  =
{{\partial {\cal H}_t}\over {\partial {\cal J}}} 
  \left( {\cal J}_t^{\tau } \right)\in   q\left(  M\right) $
corresponding to Hamiltonian
$p^*H_t \left( \eta \right) (x) =
H^{T^*M}_t\left( x, p \left( \eta \right) \right) $,
equations (\ref{EM1}) and (\ref{EM2}) of motion
becomes Lie-Poisson equation  (consult \cite{M&R&W}):
\begin{equation}\label{Hq}
{{\partial  {\cal J} ^{\tau }_t}\over {\partial t}} 
= ad^*_{\hat H_t ^{\tau }     }{\cal J}^{\tau } _t   .
\end{equation}

Classical mechanics
requires the local 
dependence on
the momentum for functionals,
while quantum mechanics
needs the wider class
of the functions
that depend on their derivatives.
For the derivative operator $D= \hbar dx^{j}\partial 
/\partial x^{j}$,
the space of the classical
functionals
and
that of the quantum functionals
are defined as
\begin{eqnarray}
C_{cl}\left( \Gamma  \right) 
&=& \left\{ p^*F 
\ \left\vert \
 p^*F  \left( \eta \right) (x) = 
{\bf F }\left( x ,    p (\eta ) (x)  \right) \
\right.
\right\} \\
C_{q\ }\left( \Gamma  \right) 
&=& \left\{   p^*F 
\ \left\vert \
 p^*F  \left( \eta \right) (x) = 
{\bf F }\left( x , p (\eta ) (x),  
..., D^{n} p (\eta ) (x) , ... \right) \ 
\right.
\right\} ,
\end{eqnarray}
and related with each other as
\begin{equation}\label{increasing}
C_{cl}\left( \Gamma   \right) 
\subset
C_q\left( \Gamma  \right)  .
\end{equation}
In other words,
the classical-limit
indicates
the limit of $\hbar \to 0$ with fixing $\vert p(\eta )(x) 
\vert $
finite at every $x\in M$, or what
the characteristic length $[x]$ and momentum $[p]$
such that $x/[x] \approx 1 $ and $p/[p] \approx 1 $
satisfies 
\begin{equation}
[p]^{-n-1} D^n p(\eta )(x)  \ll 1 .
\end{equation}
In this way, the protomechanics
realizes the {\it analyticity} of
the exact classical-limit.
The dual spaces 
make an decreasing series
of subsets:
\begin{equation}\label{decreasing}
C_{cl}\left( \Gamma    \right) ^*
\supset
C_q\left( \Gamma   \right) ^*.
\end{equation}
Thus,
quantum mechanics allows
more restricted class of 
the emergence measures
such as the density matrices
for discrete eigen wave-functions
than classical mechanics,
while it has considerably wider class of observables.

The present theory 
 also explains how  protomechanics
deduces classical mechanics and quantum mechanics,
respectively.
They will consider the space of the
synchronicities such that
\begin{equation}
\Gamma^A_{ k}  = \left\{ 
\eta \  \left\vert
\  p_j\left( \eta
\right) (\bar x) = \hbar^{A} k_j \in {\bf R} \right.
\ \ at \ a \ fixed \ point \ \bar x \
\right\} ,
\end{equation}
which requires $A=0 $ and $A=1 $ 
for
 classical case and quantum case, respectively.
The choice of reference point $\bar x$ does not
affect the deduced mechanics.
A Lagrange foliation
$\bar p$ in $TM$ further
has a
synchronicity $\bar\eta [k]\in \Gamma^A_{ k} $:
\begin{equation}
\bar p[k]=p\left( \bar  \eta [k] \right) ;
\end{equation}
and it separates every
synchronicity $\eta [k]\in \Gamma^A_{ k} $
into two parts:
\begin{equation}
\eta [k]= \bar  \eta [k] \cdot \xi .
\end{equation}
where $\xi \in \Gamma^A_{ 0} $.
Compress all the infinite 
information
of back ground  $\xi $ finally
produces classical mechanics
and quantum mechanics.
In the classical-limit,
Lie-Poisson equation (\ref{Hq}) 
deduces
the classical Liouville equation
for the induced probability density
function  $\rho_t^{T^*M} $
on cotangent space $T^*M$:
\begin{equation}\label{equation for rho}
{{\partial } \over {\partial t }}\rho^{T^*M}_t  =
\{ \rho^{T^*M}_t , H^{T^*M} \} .
\end{equation}
For canonical Hamiltonians,
Lie-Poisson equation  (\ref{Hq})
deduces the following quantum Liouville equation
for the 
density matrix $\hat \rho_t$ and the corresponding 
Hamiltonian operator
$\hat {\bf H}$:
\begin{eqnarray}\label{op}
{\partial \over {\partial t}} \hat \rho  _t  
= \left[   \hat \rho  _t  ,
\hat {\bf H} \right] /(-i \hbar ) .
\end{eqnarray}
If the Hamiltonian 
is not canonical and has the operator-ordering problem,
it will not be expressed in the summation of finite 
numbers of
polynomials of
position observable $\hat x$ and momentum observable 
$\hat p$
in general \cite{A&M}. Even so, the protomechanics
has no such trouble for the concrete
calculations
on the level of expression (\ref{Hq})
before deducing operator expression (\ref{op}).
In addition,
the present theory
proves valid also  for the half-spin
of a particle as a rigid spherical rotor
in a well-known way \cite{Holland}, while
it may also enable the conventional geometric 
interpretation
of a spinor since such an observable is not directly 
measured
as discussed later.

The present theory shares a modal interpretation
with the de Broglie-Bohm theory \cite{Holland}.
A prototypical experiment 
would always substitute
the measurement of the {\it position} of an object $Y$
not only for that of the position itself
but also for that of an observable $\hat {\bf F}$ as
 the spin, the momentum, or the energy  representing
the state of
another object $X$.
Object $Y$ can be a prepared particle to be scattered,
a radiated particle like a photon
or a classical object such as a detector's pointer,
while object $X$ is  another particle but $Y$ or
an internal freedom of a particle $Y$
for the observation of the spin of $Y$.
The following three processes
constitute such an experiment: 
\begin{enumerate}
\item the preparing process
to select an appropriate initial state for $Y$,
\item
the translating process to decompose a spectrum 
 of $Y$,
and
\item
 the detecting process to detect a particle $Y$
but not $X$.
\end{enumerate}
On the first stage of preparation,
let us suppose
that observable $\hat {\bf F}$ of
a particle or a field $X$
 has discrete igen vectors $\vert X; j \rangle $ 
such that $\hat {\bf F}
\vert X; j \rangle = j \vert X; j \rangle $ for every discrete
igen values $j$.
The initial wave function would be
prepared as $\vert \psi^{in} \rangle 
= \sum_{j}c_j\vert X;j \rangle \otimes 
\vert  Y; \phi \rangle  $
for a wave vector $\vert Y; \phi \rangle  $
of object $Y$ whose emergence frequency 
is positive everywhere: 
\begin{equation}\label{preparation}
f_Y  \left( \eta \right) (x) \geq 0.
\end{equation}
This relation requires the positivity
of the Wigner function corresponding to vector $\vert  Y; 
\phi \rangle  $:
\begin{equation}
\int_{{\bf R}^3}d^3k  \left\langle Y; \phi \left\vert 
k-{k^{\prime }\over 2}  \right.
\right\rangle 
e^{ik^{\prime }\cdot x}
\left\langle \left.
k+{k^{\prime }\over 2} \right\vert  Y; \phi \right\rangle 
\geq 0.
\end{equation}
Initial wave function $\left\vert \psi^{in} \right\rangle  $, 
on the second stage,
will be changed through the spectral decomposition into
$\left\vert \psi^{out} \right\rangle 
= \sum_{j}c_j\vert X; j \rangle \otimes 
\vert  Y; \phi_j \rangle  $,
where 
$\vert Y; \phi_j \rangle $ represents
the spatial wave function of $Y$ moving toward the j-th 
detector.
On the third stage,
the wave-reduction occurs as
the decoherence that
the density matrix loses its
nonorthogonal parts 
after the interaction with 
the measuring apparatus and/or its environment:
\begin{eqnarray}
\label{MN}
\hat \rho^{out} &=&\sum_{j,k}  c_j  c_k^*
\vert X;j \rangle \langle X; k \vert \otimes 
\vert Y;\phi_j \rangle \langle Y;\phi_ k \vert\\
 \  \ \ \ \ &\to & \ \ \ \ \ \hat \rho^{f} =\sum_{j}  c_j  c_j^*
\vert X;j \rangle \langle X; j \vert \otimes 
\vert Y;\phi_j \rangle \langle Y;\phi_ j \vert .
\end{eqnarray}
To realize decoherence (\ref{MN}),
Machida
and Namiki \cite{M&N}
considered that a macroscopic device 
is an open system that interacts
with the external environment,
and
describes the
state
of the measuring 
apparatus 
by introducing the
continuous super-selection rules
for 
Hilbert spaces.
The state of the j-th detector  
is described for
continuous measure $P$ 
on the region $L\subset M$ occupied
with considerable number of atoms
constituting the detector:
\begin{equation}
 \hat \rho^{(j)}   = 
\int_{L } d P (l ) \
\hat \rho (l ) ^{(j)}
\in \Omega^M .
\end{equation}
They further utilized the Riemann-Lebesgue
Lemma to induce
the decoherence of the density matrix
 $ \hat \rho^I $
or makes all the
off-diagonal
part zero through the interaction
between the particle and the detector.
The present theory does not only allows
the
continuous super-selection rules
but also
justifies
the utilized
approximation or limiting process
that takes
the particle number consisting
the detector as infinite,
 without serious problems
 of the objectification \cite{Espagnat}
since objects always have their own reality.
The relative frequency
that particle $Y$ appears in 
the j-th detector
should be proportional to 
the coefficient $\vert c_j\vert^2 $
since the
emergence measure of $X$ is conserved through the 
experiment
since object $X$ itself is not measured.

As discussed so far,
the present theory is a good candidate
to solve all the remained problems in quantum 
mechanics.
It meets our ordinary feeling that
a celebrated Sch\"odinger cat must know himself that
he is alive if so. 
It also revises the nonconstractive
idea that the fundamental theory must be valid 
independently
of the describing scale,
and that classical mechanics can share
an ontology with quantum mechanics.
It is expected to
provide the mathematical basis
for the mechanics in the intermediate region
between classical scale and quantum scale,
or applied to the quantum phenomena of a gravitational
field. In addition, the proposed mechanism discretizing 
the ordinary
time may be utilized to calculate the 
nonperturbative effects of the quantum field theories.


\begin{thebibliography}{99} 

\bibitem{Groenwald} H.J. Groenwald, Physica 12 (1946) 
405.

\bibitem{van Hove} van Hove,
Mem. Acad. Roy. Belg. 26 (1951)  61.

\bibitem{Maslov} V.P. Maslov and M.V. Fedoriuk, {\it
Semi-Classical Approximation 
in Quantum Mechanics} (Reidel, Dordrecht, 1981).

\bibitem{Truman} A. Truman, J. Math. Phys. 17 (1976) 
1852.

\bibitem{Pauli&Villars}
W. Pauli and F. Villars, Rev. Mod. Phys. 21 (1949) 434.

\bibitem{tHooft&Vertman}
G. 't Hooft and M. Vertman, Nucl. Phys. B 44 (1972) 
189.

\bibitem{Mittelstaedt} P. Mittelstaedt, 
{\it The Interpretation
of Quantum Mechanics and the Measurement
Process} (Cambridge University Press, 1998).

\bibitem{BLM} Paul Busch, Pekka J. Lahti adn Peter 
Mittelstaedt,
{\it The Quantum Theory of Measurement} 
(Springer-Verlag, Berlin Heidelberg,
 1994),
second edition.

\bibitem{EPR}
A. Einstein, B. Podolsky, and N. Rosen, Phys. Rev. 47 
(1935)  777.

\bibitem{Bell}
J.S. Bell, Physics 1  (1964) 195;
{\it Speakable and unspeakable in
quantum mechanics}
(Cambridge University Press, 1987).

\bibitem{Aspect}
A. Aspect, P. Grangier, and G. Roger, Phys. Rev. Lett. 49 
(1982)  91.

\bibitem{Ono} T. Ono,
Phys. Lett. A 230 (1997) 253; Phys. Lett. A 233 (1997) 
493;
the doctoral dissertation in University of Tokyo (1997).

\bibitem{I} T. Ono, 
submitted to {\it Found. Phys.} (1999); 
quant-ph/9909025.

\bibitem{Whitehead}   A.N. Whitehead, 
{\it Process and Reality}, 
edited by D.R. Griffin and D.W. Sherburne 
(The Free Press, New York and London, 1979).

\bibitem{Moyal} J.E. Moyal, 
Proc. Camb. Phil. Soc. 
45  (1949)  99.

\bibitem{M&R&W} J. Marsden, T. Ratiu, and A.
Weinstein, Trans. Am. Math. Soc. 281  (1984)  147.

\bibitem{A&M} R. Abraham and J. Marsden, {\it 
Foundation of
Mechanics}
(Addison-Wesley, Reading,
MA, 1978), second edition.

\bibitem{Holland} P.R. Holland, {\it The Quantum
Theory of Motion}
 (Cambridge University Press, 1993).

\bibitem{Wigner} E. Wigner, 
Phys. Rev. 
40 (1932) 749.

\bibitem{M&N} S. Machida and M. Namiki,
Prog. Theor. Phys.
63 (1980)  1457; Prog. Theor. Phys.
63 (1980)  1833.

\bibitem{Espagnat} 
B. d'Espagnat,
{\it Reality and the Physicist}
(Cambridge University Press, 1989),
 Addendum.


\end{thebibliography}
\end{document}